# Desarrollo y evaluación de uso de encuesta virtual de filtro sanitario en un sistema privado de educación básica de Baja California México durante contingencia sanitaria de SARS-CoV-2 (COVID-19)

## Development and evaluation of the use of a virtual health filter survey in a private primary education system in Baja California Mexico during the SARS-CoV-2 contingency (COVID-19)

Gerardo S. Romo-Cárdenas[1], María de los Ángeles Cosío-León[2], Gener José Avilés-Rodríguez[1], Nahum Isiodoro Carballo[3], Erika Zúñiga-Violante[3], Juan de Dios Sánche-López[1], Juan Iván Nieto-Hipólito[1]

[1] Facultad de ingeniería Arquitectura y Diseño, Universidad Autónoma de Baja California, Campus Ensenada, Ensenada, México.
[2] División de Posgrado e Investigación, Universidad Politécnica de Pachuca, Zempoala, México
[2] Universidad de Montemorelos, Montemorelos, México.

romo.gerardo@uabc.edu.mx, jddios@uabc.edu.mx

## Resumen

Este trabajo reporta sobre el diseño, implementación y evaluación de una encuesta virtual, utilizada como filtro sanitario remoto, aplicado en el Sistema Educativo Adventista de Baja California, durante la contingencia sanitaria generada por el virus SARS-CoV-2. Esto a fin de generar información en tiempo real y de forma segura, para salvaguardar la integridad de la comunidad educativa. Donde para cumplir con los requisitos de uso distribución, adquisición y manejo de información, se utilizaron recursos digitales gratuitos, de fácil acceso y distribución. Los cuales se adaptaron a la dinámica de comunicación del sistema educativo en cuestión. El filtro permitió adquirir información de salud de la comunidad estudiantil y facilitó la toma de decisiones respecto a la asistencia a clases por parte de los tutores, considerando las condiciones establecidas por las autoridades. A pesar de la premura en la implementación, la comunicación del sistema educativo permitió la efectiva distribución, aplicación y seguimiento del instrumento. Estrategias como estas pueden ser utilizadas para seguimiento durante la reactivación posterior a confinamientos y también analizar componentes sociales que puedan estar implicadas en estos escenarios.

*Palabras clave*: COVID-19; Encuesta de filtro sanitario; Estrategia de mitigación en contexto escolar

## Abstract

This work reports on the design, implementation and evaluation of a survey used as a remote sanitary filter, which was applied in the Adventist Educational System of Baja California. During the SARS-CoV-2 sanitary contingency. This with the intention of acquiring information safely and in real time. Allowing to safeguard the integrity of the educational community as a whole. For this, free, accessible and easily distributed digital resources were used in order to meet the necessary requirements of distribution, acquisition and management of the information. These were adapted to the previously established communication dynamic of the educational system. The use of the survey as a sanitary filter allowed to acquire relevant health information from the student community and facilitated decision-making regarding class attendance by tutors. Considering the conditions established by the authorities. Despite the rush in implementation, the communication of the educational system allowed the effective distribution, application and monitoring of the digital survey. Similar strategies can be used in the follow-up during the reactivation after confinement and to analyze social components that may be involved in these scenarios

.*Keywords*: COVID-19; Health surveillance; Mitigation strategy in school context

## 1. Introducción

Actualmente el mundo enfrenta una de las peores contingencias sanitarias de la historia reciente. La Organización Mundial de la Salud (OMS) en su sesión del 11 de marzo del 2020 declaró al brote de infección por el nuevo coronavirus llamado síndrome respiratorio grave agudo 2 (Severe Acute Respiratory Syndrome Coronavirus 2, SARS-CoV-2) como pandemia. Siendo ésta la primera pandemia causada por un coronavirus [1]. A esta fecha se habían reportado más de 125,000 casos. De los cuáles más de 40,000 fueron detectados en 118 países fuera de China; con aproximadamente 4,613 defunciones. La enfermedad asociada al SARS-CoV-2 se le ha llamado COVID-19 por las siglas CO de corona, VI de virus, D de la palabra en ingles de enfermedad (Disease) y 19 por su año de aparición (2019) [2]. El lugar inicial de la pandemia fue la ciudad de Wuhan, China, en la provincia de Hubei, posteriormente se extendió a otros países de Asia, Europa, África, Norteamérica y Latinoamérica.

A modo de estrategia de mitigación de propagación del virus, se ha reportado el cese de actividades y cierre de escuelas a fin de limitar la movilidad de la población y con ello la velocidad de propagación del virus. Como lo fue con el brote del Síndrome Respiratorio Agudo Severo (SRAS) en Singapur de febrero a mayo de 2003 [3, 4]. Esta estrategia también se aplicó dentro de los primeros cincuenta días de la contingencia sanitaria por el SARS-CoV-2 en China a principios del 2020 [5].

En México, los primeros casos se detectaron con fecha del 28 de febrero, correspondiendo a eventos de importación de virus [6]. Dado el crecimiento en la detección casos y en preparación ante la contingencia sanitaria, la Secretaría de Educación Pública (SEP) anunció con fecha del 14 de marzo, el Acuerdo Nacional de Suspensión de Actividades (02/03/20 DOF) [7]. Con lo que se suspenderían las clases a partir del viernes 23 de marzo en todos los niveles educativos. Esto como medida de aislamiento preventivo ante la pandemia del COVID-19, promoviendo que los estudiantes permanecieran en casa. El artículo segundo de este acuerdo, establecía que durante el periodo comprendido del 17 al 23 de marzo, para las escuelas de preescolar, primaria y secundaria, las madres y padres de familia o tutores deberán evitar llevar a sus hijas, hijos o pupilos, cuando estos presentaran algún cuadro de gripa, fiebre, tos seca, dolor de cabeza y/o cuerpo cortado.

Para esos efectos, la Comisión de Salud instalada en cada escuela, misma que funciona conforme a los lineamientos que para tal efecto se emitieron; determina lo que en su caso corresponda para evitar riesgos en los demás miembros de la comunidad escolar. Atendiendo, en todo momento, lo que indiquen las autoridades en materia de salud e instrumentos de monitoreo y control de la pandemia definidos a nivel federal y por estado.

Ante este escenario, se plantea necesario el uso de estrategias que permitan mitigar la propagación del virus en el contexto escolar, minimizando el contacto y permitir el tránsito de información para la toma de decisiones[8]. Para lo cual, el uso de herramientas tecnológicas como aplicaciones digitales y de tránsito de información a través de dispositivos móviles, permitirían generar información de manera segura, disminuyendo el contacto entre personas y permitiendo generar datos para la toma de decisiones [9, 10].

En este contexto, se colaboró con el Sistema Educativo Adventista de Baja California, para la generación de estrategias de control de la pandemia mediante filtros sanitarios remotos, soportados por la adquisición de datos en tiempo real, por medio de una encuesta virtual. La tarea de esta encuesta fue la de recabar información proporcionada por los padres o tutores, respecto al cumplimiento por los estudiantes de las medidas de higiene recomendadas por la Secretaría de Salud y el seguimiento de las instrucciones emitidas por la Secretaría de Educación Pública. Considerando como directriz principal la reducción del riesgo de contagio al evitar las interacciones con los diferentes miembros de la comunidad escolar.

.

Desarrollo y evaluación de uso de encuesta virtual de filtro sanitario en un sistema privado de educación básica de Baja California durante contingencia sanitaria de SARS-CoV-2 (Covid -19)3

Tabla 1. Distribución de alumnado y personal en el Sistema Educativo Adventista de Baja California

| Institución | Preescolar | Primaria | Secundaria | Preparatoria | Personal |
|---|---|---|---|---|---|
| CEH | 0 | 64 | 50 | 35 | 19 |
| COFEMO | 13 | 111 | 70 | 45 | 19 |
| CSYS | 29 | 229 | 156 | 112 | 41 |
| COFIM | 0 | 66 | 51 | 0 | 14 |
| Total | 42 | 470 | 327 | 192 | 93 |

## 1. Desarrollo de encuesta virtual de filtro sanitario para un sistema educativo privado del Estado de Baja California

El Sistema Educativo Adventista de Baja California (SEA) donde se llevó a cabo el estudio es un sistema educativo privado compuesto por cuatro instituciones distribuidas en la misma cantidad de municipios, de los cinco del estado: Colegio Elena Harmon en Mexicali (CEH), Colegio Fernando Montes de Oca en Tecate (COFEMO), Colegio Salud y Saber en Tijuana (CSYS) y Colegio Francisco I. Madero en Ensenada (COFIM). Donde laboran 93 empleados entre docentes, administrativos y personal de apoyo. Así mismo, están inscritos 1031 estudiantes entre preescolar, primaria, secundaria y preparatoria, tal y como se muestra en la Tabla 1

La Fig. 1 muestra la distribución geográfica de las escuelas donde se aplicó la encuesta de filtro sanitario en el estado de Baja California y los municipios de impacto.

A modo de estrategia general para el filtro sanitario virtual, se planteó un esquema donde cada padre de familia o tutor se le compartía un acceso a la encuesta virtual. En este instrumento se proporcionaba información de identificación del estudiante. Posteriormente se llenaban las preguntas referentes a salud, activando la operación del filtro sanitario en el momento de llenar la misma. Después, la información generada en este filtro virtual se utilizaba como información de soporte en un filtro secundario físico en la entrada de los planteles.

De forma individual, cada institución manejaría la información a fin de obtener el panorama de salud de su comunidad y administrar las decisiones de manera local. Más dado que la información se integra de forma digital, es factible integrar y hacer análisis y exploración de datos y considerar también el manejo de decisiones a nivel de sistema educativo estatal y agregar una exploración tanto epidemiológica como geo estadístico del progreso de la enfermedad [11, 12].

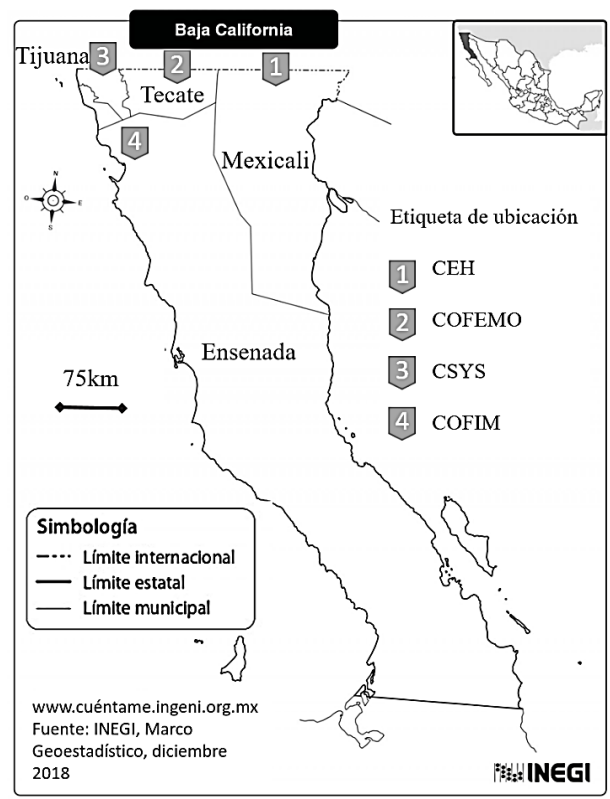

Fig.1 Área de estudio de las escuelas del Sistema Educativo Adventista del estado de Baja California.

Otra componente importante respecto al desarrollo de la encuesta de filtro es de naturaleza sanitaria. Se partió desde el supuesto de que se conoce poco respecto al virus SARS-CoV-2, siendo una variable a considerar en el diseño [11]. Inicialmente se había reportado que los pacientes con la enfermedad Covid-19, presentaron síntomas clínicos de tos seca, disnea o dificultad para respirar y fiebre [13]. Al ser

Desarrollo y evaluación de uso de encuesta virtual de filtro sanitario en un sistema privado de educación básica de Baja California durante contingencia sanitaria de SARS-CoV-2 (Covid -19)4 esta una enfermedad de índole respiratoria, los síntomas o características clínicas podrían en algunos casos coincidir con los de un resfriado o gripe.

Considerando los requerimientos que se comunicaron para el funcionamiento de los filtros sanitarios y la incertidumbre que se pudiera considerar en los padres y tutores respecto a la decisión de llevar a sus hijos a la escuela [14]. El primer factor de decisión de operación de estos filtros, se basó en preguntar si el alumno asistiría al plantel. En caso de ser afirmativa la respuesta. Se requería confirmar todos los requisitos establecidos por la autoridad, que incluía confirmar el lavado de manos, no haber presentado fiebre y no tener tos o síntomas de gripa. En caso de no poder confirmar todos estos, por medio de programación lógica, no se permitía avanzar en el llenado de la encuesta y se regresaba a preguntar si el alumno asistiría al plantel. Cuando se elegía no asistir al plantel, se requería elegir una opción de la decisión entre las que estaban: decidir hacer aislamiento preventivo, se llevaba al estudiante a alguna unidad de salud u otra. Posteriormente la información pasaba al concentrado que se manejaba a la entrada de cada plantel. El esquema del proceso de información planteado en la estrategia de la encuesta virtual de filtro sanitario se visualiza en el diagrama de la Fig 2.

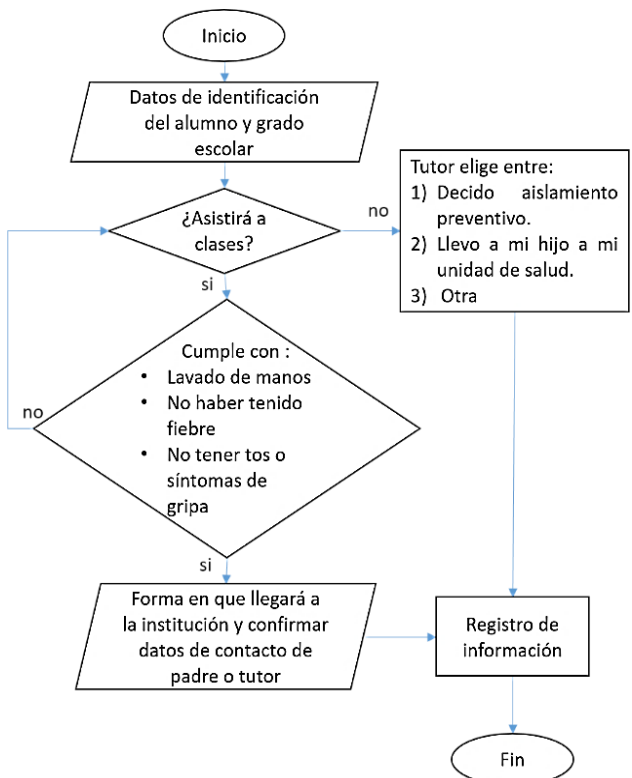

Fig. 2 Esquema de funcionamiento de filtro sanitario virtual para el Sistema Educativo Adventista.

Dado el escenario urgente requerido para la implementación de la herramienta para el filtro sanitario, la encuesta se realizó en una plataforma abierta de encuestas de Google, Google Forms ®. La cual facilitaba el diseño para cada institución, así como con el requisito de distribuir el instrumento, facilitarle a los padres o tutores la tarea de contestar, ya que se puede sea con una computadora o un teléfono móvil y también concentrar la información tan pronto como los padres lo confirmaran. Esto para su uso tanto como filtro secundario, así como para integrar reportes y análisis posterior. Así mismo, para su ágil distribución con la comunidad de padres, se consideraron los canales de comunicación de mensajería instantánea por telefonía móvil, normalmente usados entre los docentes y los padres

Con lo que se propuso el siguiente protocolo de uso:

1. Anunciar a los padres de familia de llenar el cuestionario, el cual sigue los parámetros de salud establecidos por la autoridad correspondiente.

2. Compartir por los grupos de mensajería instantánea (WhatsApp®) de cada salón, la liga de acceso del formulario y recordar a diario el llenado.

3. El padre, madre o tutor, llena por cada alumno el cuestionario previo a dirigirse al plantel y atiende dudas con el docente a cargo del grupo.

4. En el filtro sanitario de la escuela, se verificará el archivo de datos capturados para confirmar la respuesta.

5. Al final del día se hace un análisis de la información recopilada.

En la figura 3 se observa un diagrama de bloques con el flujo de información de la estrategia seguida para la aplicación de la encuesta del filtro sanitario virtual.

La estrategia de flujo de información plasmada en la figura 3 considera la necesidad de replicar el anuncio de la aplicación del filtro sanitario y de la integración de la información para su análisis.

Desarrollo y evaluación de uso de encuesta virtual de filtro sanitario en un sistema privado de educación básica de Baja California durante contingencia sanitaria de SARS-CoV-2 (Covid -19)5

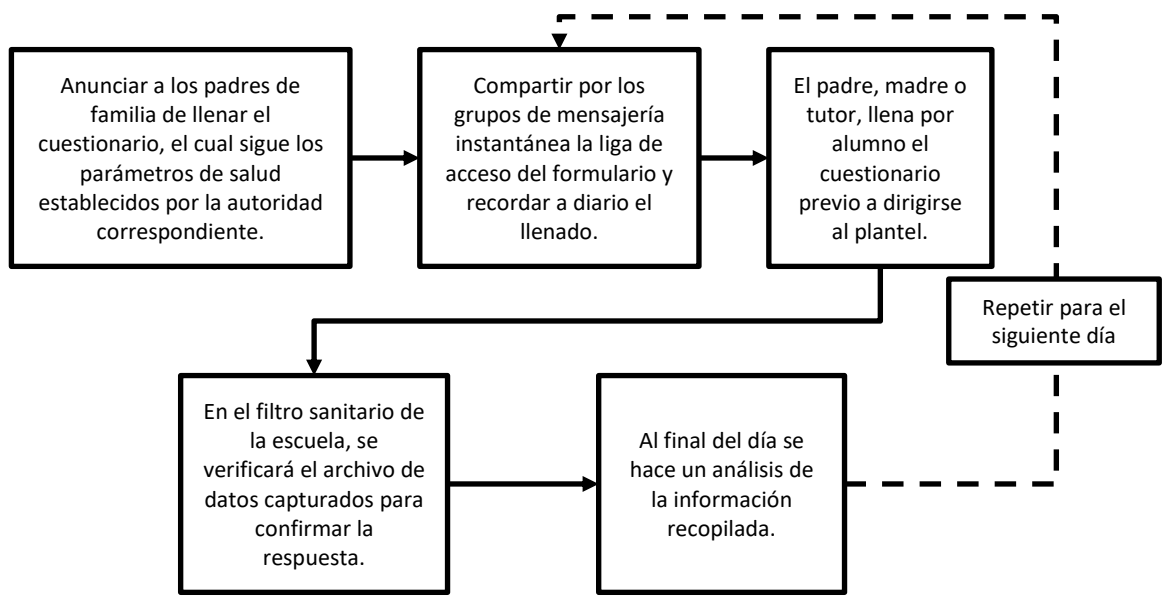

Fig. 3. Diagrama de flujo de información de encuesta virtual de filtro sanitario durante periodo de contingencia.

Posterior a la elaboración y prueba el instrumento virtual; se revisó que este protocolo con las autoridades del SEA y de los planteles, para ser comunicado posteriormente a los docentes, quienes se encargaron de comunicar a través de memorándums oficiales a los padres de familia, con las instrucciones y ligas de acceso a los cuestionarios del filtro sanitario virtual.

Como se mencionó en el protocolo de uso y en la figura 3. Se tomó ventaja de los canales de comunicación de mensajería instantánea previamente usados entre docentes y padres de familia para divulgar la información relevante a la encuesta de filtro virtual entre las familias de los estudiantes. Del mismo modo este canal sirvió para poder atender las dudas asociadas al llenado o relevantes a las respuestas que darían los padres en la encuesta.

## 2. Aplicación de encuesta virtual de filtro sanitario para el Sistema Educativo Adventista de Baja California

La implementación se inició a partir del 17 de marzo de 2020, en los 4 planteles, siguiendo el protocolo de uso propuesto y con una participación del 91% de la población estudiantil. Al recopilar los datos de forma remota, permitió que la información se integrara en tiempo real. A la llegada de los estudiantes, se verificó el haber recibido la información por parte de la comunidad estudiantil en la entrada de cada plantel, funcionando a su vez como filtro secundario.

Sin embargo, dado el registro de casos positivos de Covid-19 en la localidad, la autoridad Estatal decidió y anunció la suspensión de actividades académicas a partir del 18 de marzo [15].

Dentro de los resultados del día en que se aplicó, se observó que en 73 casos, el padre o tutor reportó que el alumno/a no se presentaría en el plantel ese día. La distribución de frecuencias en las respuestas de este caso, se observan en la Tabla 2.

Tabla 2. Distribución de respuestas en estudiantes que reportaron no asistir al plantel.

| Institución | Decido hacer resguardo preventivo | Llevo a mi hijo/a a mi unidad de salud | Otra |
|---|---|---|---|
| CEH | 23 | 7 | 4 |
| COFEMO | 5 | 1 | 3 |
| CSYS | 6 | 0 | 8 |
| COFIM | 9 | 0 | 7 |

En la Tabla 2 se puede observar que, en 43 casos, el padre o tutor decidió no llevar al estudiante al plantel decidiendo por hacer un resguardo preventivo. En 8 ocasiones declararon llevar al alumno a una unidad de salud y en 22 ocasiones decidieron no llevar al alumno a clases por otras razones.

De las interacciones entre docentes y padres de familia, derivadas del funcionamiento y programación lógica de la encuesta, en 16 ocasiones se les sugirió a los padres de familia no llevar a su hijo/a al plantel dado que no se cumplía con los requisitos de salud necesarios. En la totalidad de estos se debió a que el alumno presentaba algún síntoma asociado a una patología respiratoria.

En el instrumento de filtro, también se hacía seguimiento al modo en que el estudiante llegaría al plantel. Así como confirmar teléfono de contacto.

Con la información recopilada, las instituciones asociadas al Sistema Educativo Adventista, pudieron entregar según fue requerido, los reportes a las autoridades correspondientes.

## 3. Percepción del uso de encuesta virtual de filtro sanitario por parte del personal del Sistema Educativo Adventista de Baja California

En una fecha posterior a la aplicación y a modo de seguimiento, se realizó una encuesta anónima al personal de los planteles, para revisar desde su perspectiva, la efectividad del uso de la encuesta como filtro sanitario.

En la tabla 3 se observa la distribución de respuestas de la encuesta de seguimiento. Donde al personal se preguntaba su opinión respecto a la facilidad de distribución y uso, la información relevante a la salud de la comunidad estudiantil y su percepción sobre la efectividad general del filtro

Tabla 3. Distribución de respuesta en porcentajes en encuesta de seguimiento a personal del Sistema Educativo Adventista de Baja California, respecto a uso de encuesta virtual como filtro sanitario.

| Pregunta | Total desacuerdo | | | | Total acuerdo |
|---|---|---|---|---|---|
| | 1 | 2 | 3 | 4 | 5 |
| Tenía claro la intensión del cuestionario de filtro sanitario | - | - | 3.37 | 17.98 | 78.65 |
| Considero que fue fácil distribuir el cuestionario con la comunidad académica | - | - | 5.98 | 34.02 | 60 |
| El cuestionario de filtro sanitario permitió informar a la comunidad académica sobre la situación de la contingencia sanitaria. | - | - | 4 | 34.78 | 61.22 |
| | Falso | | | Cierto | |
| La aplicación del cuestionario de filtro sanitario me permitió interactuar con los padres respecto al estado de salud de sus hijos. | 8.77 | | | 91.23 | |
| Considero que la aplicación del cuestionario tuvo una función efectiva como filtro sanitario. | 18.3 | | | 81.7 | |
| | No | | | Si | |
| La aplicación del cuestionario de filtro sanitario me permitió sugerir / indicar al padre de familia a aplicar resguardo preventivo con al menos un estudiante | 5 | | | 95 | |

De observar la misma tabla 3, sobresale el hecho de que 91.23% del personal consideró que les permitió interactuar con los padres o tutores respecto al estado de salud de sus hijos. Respecto a la claridad de la estrategia, la facilidad de distribución y el manejo de información, más del 90% de los participantes reportaron estar de acuerdo dentro de los dos niveles para estos parámetros. Finalmente, el 95% de los encuestados reportó sentirse en posibilidad para poder sugerir o indicar al padre de familia, resguardo preventivo para su hija/o.

## Conclusiones

La contingencia sanitaria derivada del SARS-CoV-2, ha requerido de la planeación e implementación de respuestas urgentes para atender diversas situaciones. En este trabajo se reportó sobre el diseño, implementación y evaluación de una encuesta virtual de filtro sanitario, a fin de salvaguardar la integridad de la comunidad educativa. Este desarrollo se diseñó y elaboró de forma expedita, donde para cumplir con los requisitos necesarios, se utilizaron recursos digitales gratuitos y de fácil acceso y distribución, para ser adaptados a la dinámica del sistema educativo en cuestión. El diseño lógico de la encuesta permitió realizar un seguimiento digital de la información de salud de la comunidad estudiantil de un sistema educativo a nivel estatal. Permitió a los docentes y personal académico el filtrar posibles casos de estudiantes con alguna patología respiratoria. Apoyando a las autoridades del sistema educativo en cuestión, en poder adquirir información de forma remota, disminuyendo el riesgo entre el personal y los estudiantes. Así mismo, se promovió la comunicación respecto a temas de salud oportunos, dentro de toda la comunidad escolar.

Tomando en cuenta el alcance de cobertura en la población del SEA, la información de proveyó, los diálogos entre todos los actores del mismo y las decisiones que permitió tomar a las autoridades académica; se puede considerar que el instrumento tuvo un manejo eficiente, aunado también a que este se distribuyó ágilmente en todos los planteles del sistema educativo estudiado. Permitiendo así generar información para la toma de decisiones que podrían tener un efecto en la mitigación de la pandemia en el contexto escolar. Cabe mencionar que para esto la aceptación del uso del filtro en la comunidad académica para la generación y flujo de información fue de suma importancia. Esto posiblemente se logró debido a la dinámica de comunicación previamente establecida dentro de la comunidad del Sistema Educativo Adventista de Baja California. El cual se observó tanto entre administradores, directivos y docentes en la etapa de diseño y preparación. Así como entre padres de familia o tutores con los docentes encargados de grupo, en la etapa de implementación y adquisición de datos.

Al momento de redactar este trabajo, el Sistema Educación Adventista del Norte de México planeaba replicar esta misma estrategia para las 32 escuelas ubicadas en los estados del norte del país, para el seguimiento sanitario en la etapa de reactivación.

Este tipo de estrategias, que permiten la interacción, seguimiento y adquisición de información de salud, pueden ser implementadas también en los procesos de reactivación de actividades, a fin de fomentar el seguimiento y toma de decisiones adecuados para diversas comunidades. Así mismo, con el uso de tecnologías como las presentadas en este trabajo, sería posible adquirir información de índole social que permita estudiar y comprender fenómenos implicados a situaciones de riesgo sanitario a nivel de comunidad.

## Agradecimientos

## Referencias

[1] I. Adhanom, "WHO Director-General´s opening remarks at the Mission briefing on COVID-19-11 March 2020," ed. Ginebra, Suiza: World Health Organization, 2020.

[2] O. Vega-Vega, M. Arvizu-Hernández, J. G. Domínguez-Cherit, J. Sierra-Madero, and R. Correa-Rotter, "Prevención y control de la infección por coronavirus SARS-CoV-2 (Covid-19) en unidades de hemodiálisis," *Salud Pública de México,* 2020.

[3] K. P. Chan, "Control of severe acute respiratory syndrome in singapore," *Environmental health and preventive medicine,* vol. 10, pp. 255-259, 2005.

[4] C.-C. Tan, "SARS in Singapore-key lessons from an epidemic," *Annals-Academy of Medicine Singapore,* vol. 35, p. 345, 2006.

[5] H. Tian, Y. Liu, Y. Li, C.-H. Wu, B. Chen, M. U. Kraemer*, et al.*, "The impact of transmission control measures during the first 50 days of the COVID-19 epidemic in China," *MedRxiv,* 2020.

[6] S. d. Salud. (2020, 20/04/2020). *Información referente a casos de Covid-19 en México.* Available:

Desarrollo y evaluación de uso de encuesta virtual de filtro sanitario en un sistema privado de educación básica de Baja California durante contingencia sanitaria de SARS-CoV-2 (Covid -19)8

[7] S. d. Gobernación, "Acuerdo 02/03/20 Sobre suspensión de actividades en todos los niveles educativos y dependientes de la Secretaría de Educación Pública.," in *Diario Oficial de la Federación*, ed. Ciudad de Mexico: Secretaria de Gobernación, 2020.
[8] A. Brito and S. M. Garza, "Escuelas sin COVID: Estrategias y medidas sanitarias para minimizar el riesgo de contagio en las escuelas."
[9] N. Ahmed, R. A. Michelin, W. Xue, S. Ruj, R. Malaney, S. S. Kanhere*, et al.*, "A survey of covid-19 contact tracing apps," *IEEE Access,* vol. 8, pp. 134577-134601, 2020.
[10] N. Oliver, B. Lepri, H. Sterly, R. Lambiotte, S. Deletaille, M. De Nadai*, et al.*, "Mobile phone data for informing public health actions across the COVID-19 pandemic life cycle," ed: American Association for the Advancement of Science, 2020.
[11] S. K. Dey, M. M. Rahman, U. R. Siddiqi, and A. Howlader, "Analyzing the epidemiological outbreak of COVID-19: A visual exploratory data analysis approach," *Journal of medical virology,* vol. 92, pp. 632-638, 2020.
[12] C. Zhou, F. Su, T. Pei, A. Zhang, Y. Du, B. Luo*, et al.*, "COVID-19: Challenges to GIS with big data," *Geography and Sustainability,* 2020.
[13] H. Lu, C. W. Stratton, and Y. W. Tang, "Outbreak of Pneumonia of Unknown Etiology in Wuhan China: the Mystery and the Miracle," *Journal of Medical Virology,* 2020.
[14] N. Chater, "Facing up to the uncertainties of COVID-19," *Nature Human Behaviour,* pp. 1-1, 2020.
[15] G. d. B. California, "Aviso de suspensión de clases en el estado de Baja California por contingencia Covid-19," D. O. d. B. California, Ed., ed. Mexicali, Baja California: Diario Oficial de Baja California, 2020.

https://datos.gob.mx/busca/dataset/informacion-referente-a-casos-covid-19-en-mexico

*Información de Contacto de los Autores*:

**Gerardo S. Romo-Cárdenas**
UABC Ensenada, Carr. Transpeninsular 3917 , 22860
Ensenada
México
romo.gerardo@uabc.edu.mx

**Ma. De los Ángeles Cosío-León**
Carretera Ciudad Sahagún-Pachuca Km. 20, Ex-Hacienda de Santa Bárbara, 43830
Pachuca
México
ma.cosio.leon@upp.edu.mx

**Gener J. Avilés Rodriguez**
UABC Ensenada, Carr. Transpeninsular 3917 , 22860
Ensenada
México
gener.aviles@uabc.edu.mx

**Nahum Isiodoro Carballo**
Sistema Educativo Adventista Libertad 130, 21400
Tecate,
México
umnsea1@gmail.com

**Erika Zúñiga Violante**
Sistema Educativo Adventista Libertad 130, 21400
Tecate,
México
erikaz@um.edu.mx

**Juan de Dios Sánchez López**
UABC Ensenada, Carr. Transpeninsular 3917 , 22860
Ensenada
México
jddios@uabc.edu.mx

**Juan Iván Nieto Hipólito**
UABC Ensenada, Carr. Transpeninsular 3917 , 22860
Ensenada
México
jnieto@uabc.edu.mx

**Dr. Gerardo S. Romo Cárdenas**

Ingeniero Físico Industrial por el Tecnológico de Monterrey, M. C. en Óptica por el CICESE. Doctor en Ciencias por la UABC. Actualmente es profesor investigador en el programa de Bioingeniería de la Universidad Autónoma de Baja California.

**Dra. María de los Ángeles Cosío León**
Realizó estudios de Maestría en Computación con especialidad en algoritmos para Redes y Conectividad por la Universidad de Colima, México. Obtuvo el grado de Doctora en Ciencias por la Universidad Autónoma de Baja California Campus Ensenada, México. Actualmente es profesora investigadora en la Universidad Politécnica de Pachuca

**Dr. Gener Avilés Rodríguez**

Desarrollo y evaluación de uso de encuesta virtual de filtro sanitario en un sistema privado de educación básica de Baja California durante contingencia sanitaria de SARS-CoV-2 (Covid -19)9

Médico Cirujano y Partero por la Universidad de Montemorelos. Maestro en Ciencias por la Universidad Autónoma de Baja California. Actualmente estudiante de Doctorado en Ciencias en la Universidad Autónoma de Baja California.

**MC. Nahum Isidoro Carballo**
Doctorado en Educación, Maestría en Educación Administrativa, Licenciatura en Educación Área Quimica y Biologia. 26 años de experiencia administrativa y docente en educación básica, media superior y superior.

**Dra. Erika Zúñiga Violante**
Doctora en Medio Ambiente y desarrollo sustentable, Maestría en Manejo de Ecosistemas de Zonas áridas, Ingeniera Química. 16 años de experiencia docente en educación básica y superior.

**Dr. Juan de Dios Sánchez López**
Recibió la licenciatura en Ingeniería eléctrica de la Facultad mecánica y eléctrica del Instituto Tecnológico de la ciudad de Madero. Recibió su M.C. y Doctorado del Centro de Investigación CICESE, México Desde agosto de 1993 es profesor titular en la Universidad Autónoma de Baja California (UABC), México.

**Dr. Juan Iván Nieto Hipólito**
Recibió su grado de M.C. de CICESE en 1994. Doctorado por el Departamento de Arquitectura de Computadores en el Politécnico Universidad de Cataluña (UPC, España) en 2005. Desde agosto de 1994 es profesor titular en la Universidad Autónoma de Baja California (UABC, México).